# Rosette spectroscopic imaging for whole-brain metabolite mapping at 7T: acceleration potential and reproducibility

Zhiwei Huang[1,2], Uzay Emir[4,5,6], André Döring[1,2], Antoine Klauser[7], Ying Xiao[1,2,3], Mark Widmaier[1,2,3], Lijing Xin[1,2,3]

1. Center for Biomedical Imaging (CIBM), Switzerland
2. Animal Imaging and Technology, Ecole Polytechnique Fédérale de Lausanne (EPFL), Lausanne, Switzerland
3. Institute of Physics (IPHYS), Ecole Polytechnique Fédérale de Lausanne (EPFL), Lausanne, Switzerland
4. University of North Carolina at Chapell Hill, Department of Radiology
5. University of North Carolina at Chapell Hill, Biomedical Research Imaging Center (BRIC)
6. University of North Carolina at Chapell Hill, Joint Department of Biomedical Engineering
7. Advanced Clinical Imaging Technology, Siemens Healthineers International AG, Lausanne, Switzerland

*Corresponding author:

Lijing Xin, Ph.D.

EPFL AVP CP CIBM-AIT

Station 6, CH-1015 Lausanne, Switzerland

lijing.xin@epfl.ch

Tel.: + 41 21 693 0597

**Word count: 4333**

# Abstract

Whole-brain proton magnetic resonance spectroscopic imaging ($^1$H-MRSI) is a non-invasive technique for assessing neurochemical distribution in the brain, offering valuable insights into brain functions and neural diseases. It greatly benefits from the improved SNR at ultrahigh field strengths ($\geq$7T). However, $^1$H-MRSI still faces several challenges, such as long acquisition time and severe signal contaminations from water and lipids. In this study, 2D and 3D short TR/TE $^1$H-FID-MRSI sequences using rosette trajectories were developed with spatial resolutions of 4.48×4.48 mm² and 4.48×4.48×4.50 mm³, respectively. Water signals were suppressed using an optimized Five-variable-Angle-gaussian-pulses-with-ShorT-total-duration of 76 ms (FAST) water suppression scheme, and lipid signals were removed using the $L_2$ regularization method. Metabolic maps of major $^1$H metabolites were obtained within 5:40 min with 16 averages and 1 average for the 2D and 3D acquisitions, respectively. Excellent inter-session reproducibility was shown, with the coefficients of variance (CV) being lower than 6% for N-Acetyle-L-aspartic acid (NAA), Glutamate (Glu), Choline Chloride and glycerophosphocholine (tCho), Creatine and Phosphocreatine (tCr), and Glycine and Myo-inositol (Gly+Ins). To explore the potential of further accelerating the acquisition, compressed sensing was applied retrospectively to the 3D datasets. The structural similarity index (SSIM) remained above 0.85 and 0.8 until R = 2 and 3 for the metabolite maps of Glu, NAA, tCr, and tCho, indicating the possibility for further reduction of acquisition time to around 2min.

# Introduction

Proton magnetic resonance spectroscopic imaging ($^1$H MRSI) is a non-invasive technique to assess the spatial distribution of metabolites in the human brain. The free induction decay (FID) MRSI sequence is the most popular among different sequences due to its high signal-to-noise ratio (SNR) using ultra-short TE and short TR. However, the long acquisition time of the conventional phase encoding scheme still limits its application in clinical settings.

Higher accelerations can be achieved with spatial-spectral encoding (SSE) techniques[1]. By sampling spatial and spectral information simultaneously utilizing the high-slew-rate gradient waveforms, SSE techniques allow 25 to 170 times faster MRSI scans than conventional phase-encoding MRSI[2]. Common SSE techniques exploit trajectories[1] like spiral[3], rosette[4], concentric rings[5,6], and echo-planar trajectories[7].

Several approaches have been proposed to further shorten the acquisition time. Parallel imaging techniques, such as SENSE, GRAPPA and CAIPIRINHA[8–10], leveraging the surplus on spatial information provided by multi-channel coil arrays, can help accelerate data collection during the acquisition stage; but the acceleration rate is limited by noise amplification due to the reduced number of phase encoding lines and the geometric factor[11]. Compressed sensing (CS) with random under-sampling patterns, leveraging the intrinsic sparsity of the signal, can surpass the Nyquist sampling criteria enabling acceleration rates of up to 3-5 for MRSI in the brain[12,13].

Ultra-high field strengths offer additional benefits, including higher SNR and improved spectral resolution, which can enhance the detection and quantification of metabolites, but they also require broader spectral bandwidth, which poses challenges due to the restriction in maximum gradient amplitudes and slew rates. One solution to alleviate the restriction is to use temporal interleaves, but it will increase the acquisition time and potentially introduce other artifacts[1]. Among different k-space sampling trajectories, the rosette trajectory exhibits higher flexibility and poses less demands on the gradients. Besides, the rosette trajectory inherently samples the k-space center more densely than the periphery, which can give improved SNR and insensitivity to motions[14,15]. Moreover, the trajectory was designed based on the stochastic trajectory[16], which is intrinsically incoherent, making it more compatible with compressed sensing for further acceleration.

The inhomogeneous $B_1^+$ distribution at high magnetic fields poses another challenge. It deteriorates the overall water suppression efficiency, leading to high water residual and substantial water sidebands overlapping with metabolite signals, making spectral fitting and metabolite quantification challenging. The established WET water suppression scheme (Water suppression Enhanced through $T_1$ effects)[17] is widely used in MRS and MRSI sequences[18] given its total duration and SAR level. However, it suffers from reduced global water suppression efficiency at 7T and above. With the incorporation of ultra-short TE and non-Cartesian trajectories which necessitate stronger gradients, the sideband problem gets worse[19]. Several studies optimized the water suppression scheme based on the WET scheme[13,20,21], but no study has shown the performance before and after the optimization.

Until now, rosette spectroscopic imaging (RSI) is mostly demonstrated at 3T, where the trajectory parameters were optimized for temporal SNR[4]. However, when it comes to the whole-brain RSI at 7T, few studies have been reported. Schirda C. et al developed RSI at 7T with LASER pre-localization and outer volume suppression (OVS) of the skull lipid signal, exciting the inner part of the brain[22]. Whole-brain MRSI including the measurements in the cortical regions often does not apply OVS, and relies on lipid removal strategies such as the $L_2$ regularized lipid removal[23], data extrapolation[24], FLIP[25] etc. A 2D whole-brain $^1$H-FID RSI at 7T with 2mm in-plane resolution with three temporal interleaves was reported with an acquisition time of around 6min[20]. Recently, a modified egg-shaped rosette trajectory with two temporal interleaves was proposed. It allows 2D measurements with 3.44mm in-plane resolution and a scan time of 6:04min, as well as 3D measurements with a resolution of 3.44x3.44x4.47 mm$^3$ and a scan time of 19min [26].

This study seeks to advance the $^1$H RSI at 7T by improving global water suppression performance, shortening TE, evaluating the reproducibility of the estimated metabolite maps, lipid removal performance and further reducing the acquisition time with CS for whole brain metabolites mapping.

## Materials and Methods

The sequence diagram is shown in Figure 1a. An asymmetric amplitude-modulated RF pulse (P10) with a duration of 2048 μs and a bandwidth of 2700 Hz was used for excitation to enable ultra-short TE of sub-milliseconds.

**Trajectory Design.** The rosette trajectory can be described by a rapid sinusoidal oscillation rotating in the $k_x - k_y$ plane[16] (1), where $\text{kmax} = \frac{N_x}{2\text{FOV}}$.

$$k(t) = k_{max} \sin(\omega_1 t) \, e^{i\omega_2 t} \quad (1)$$

The selection of $\omega_1$ and $\omega_2$ is constrained by the maximum gradient amplitude, slew rate, and spectral bandwidth (2,3,4).

$$|G_x| \leq \frac{2\pi}{\gamma} k_{max} \max(\omega_1, \omega_2) \quad (2)$$

$$|S| \leq \frac{2\pi}{\gamma} k_{max}(\omega_1^2 + \omega_2^2) \quad (3)$$

$$\gamma \text{FOV } G_{max} + SBW \leq \frac{1}{dt} \quad (4)$$

Here, $\omega_1 = \omega_2 = 2000\pi$ was chosen to achieve the maximum spatial resolution with a spectral bandwidth of 2000Hz to cover a $^1$H spectral range of 6.7ppm at 7T. The maximum gradient amplitude and slew rate were 16.46 mT/m and 206.96 mT/m/s, respectively, restricted by the peripheral nerve stimulation limits. The resulting in-plane resolution is 4.48x4.48 mm². $N_{sh}$ = 79 petals were acquired for a fully sampled 2D acquisition[4].

To improve the spatial response function, a rotating average scheme was used. Instead of $N_{sh}$ petals repeated by the number of average (NA) times, $NA \times N_{sh}$ petals were acquired (figure 1b). The spatial response functions for the normal averaging mode and rotating averaging mode are shown in Figure 1c. For the 3D acquisition, a weighted stack of rosette trajectories along $k_z$ direction with a spherical bound in 3D was used. The maximum $k_{xy}$ and number of shots $N_{sh_z}$ in each $k_z$ plane are given by (5,6):

$$k_{xy_{max}} = \max(\sqrt{k_{max}^2 - k_z^2}, 0.2 k_{max}) \quad (5)$$

$$N_{sh_z} = ceil(\frac{k_{xy_{max}}}{k_{max}} N_{sh_0}) \quad (6)$$

where $N_{sh_0}$, $N_{sh_z}$ are the number of petals in the slices with $k_z$ being 0 or z, $k_{max}$ is the maximum k value in the $k_x - k_y$ plane in the central slice. The spatial resolution is 4.48x4.48x4.50mm³, with 20 slices covering 90mm FOV along the z-axis.

**Water Suppression Scheme.** The established WET water suppression scheme[17] consists of three Gaussian pulses with flip angles of 89.2°, 83.4°, 160.8° and equal inter-pulse delays of 60ms[17] (total duration of 180 ms). The scheme was optimized to 1.5T and its efficiency is sensitive to $B_1^+$ inhomogeneity. To improve the water suppression efficiency at 7T, we optimized the water suppression scheme tailoring for $B_1^+$ insensitivity (figure 2a). The new water suppression scheme consists of Five variable Angle gaussian pulses with a ShorT total duration of 76 ms (FAST). The flip angles are 100°, 80°, 125°, 65°, 170°, and the inter-pulse delays are 15ms, 15ms, 15ms, 17ms, 14ms, respectively. The crusher gradients were set so that no spurious echoes were seen in the spectra, with the moments being 168, 168, 168, 235, 235 ms*mT/m respectively. The FAST scheme is robust to [-50, +60] % variation in the $B_1^+$ field, achieving a 98% suppression efficiently (figure 2a), which allows to reduce the minimum TR from 370ms to 270ms.

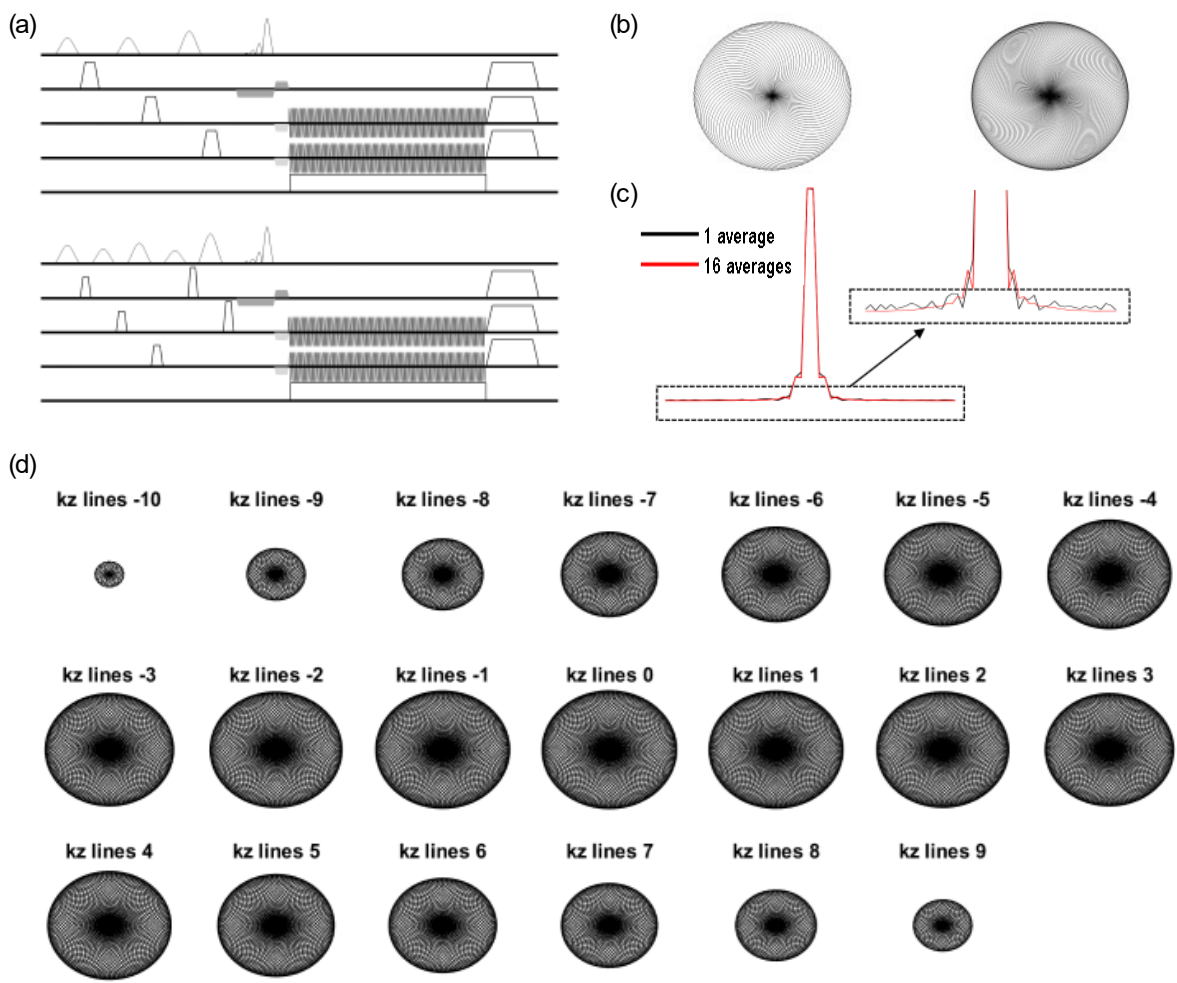

Figure 1. a) The sequence diagram: top: the WET water suppression scheme, bottom: the FAST water suppression scheme; dark gray: localization and corresponding rewinding gradients; light gray: rewinding gradients for the trajectory; blank with black outline: spoiler gradients; b) The 2D rosette trajectory with 79 petals (left, NA=1) and 100 petals (right, NA~1.27); half petals were shown for the sake of visualization, c) Comparison of normalized spatial response functions of rotating average and single average with NA=16; d) The 3D trajectory with a weighted stack of rosette trajectories.

**Phantom and participants**. The sequence was tested on a spherical phantom (*Gold Standard Phantoms, spectre,* diameter of 18cm) for validation, which contains seven metabolites at physiological pH and concentrations (12.5mM N-Acetyle-L-aspartic acid (NAA), 10.0mM Creatine (Cr), 3.0mM Choline Chloride (Cho), 7.5mM Myo-inositol (Ins), 12.5mM Glutamate (Glu), 5.0mM Lactate (Lac), and 2.0mM GABA). Three healthy subjects (28-40 years old, 2 males and 1 female) were scanned for this study. All volunteers gave informed consent in accordance with the Swiss cantonal ethics committee before the experiments.

**MR acquisition**. All experiments were performed on a 7T MAGNETOM Terra. X scanner (Siemens Healthineers, Forchheim, Germany), with an 8-channel transmit coil with 32 receiver arrays (Nova Medical, Inc. Wilmington MA, USA). The maximum allowed gradient amplitude and slew rate are 130mT/m and 250mT/m/s, respectively. First-, second- and third-order $B_0$ shimming were implemented using the vendor-pre-implemented *DESS Advanced* scheme.

Phantom experiments: 2D measurements were performed with a matrix size of 50x50 (NA=1, total scan time 30s). TE/TR = 2.32ms/370ms was used to form a comparison with the product CSI sequence. Considering that the *in vitro* linewidth is narrower than that *in vivo*, a 5 Hz exponential filter was applied in the temporal domain.

*In vivo* experiments: A $T_1$-weighted anatomical image was first acquired with the MP2RAGE sequence (TE/TR = 1.54 ms/5500 ms, $TI_1/TI_2$ = 750 ms/2350 ms, $\alpha_1/\alpha_2$ = 4°/5°, FOV=169 x 248 x 256mm$^3$, matrix size = 256 x 404 x 416)[27]. The 2D RSI data were acquired with a FOV of 224 x 224mm$^2$, a matrix size of 50x50, and a slice thickness of 10 mm. The acquisition took 20s per shot. 16 averages were acquired to improve the SNR, leading to a total acquisition time of 5:40min. An additional water scan was obtained for coil combination and metabolite quantification. The 3D RSI data were acquired with an FOV of 224x224x90mm$^3$. An excitation slab of 65mm was adopted to avoid aliasing artifacts. The matrix size was 50x50x20, leading to a nearly isotropic resolution of 4.50mm. The total acquisition was 5:30min for the metabolite data (TR=270ms), and 4:00min for the water scan (TR=190ms). The TE for the 2D and 3D acquisitions were 0.87ms and 0.83ms, respectively. The flip angle (FA) was set to the Ernst angle assuming the average T1 being 1800ms[28] for the metabolites (TR=270ms, FA=31°; TR=190ms, FA=26°). Both acquisitions were repeated twice to evaluate the inter-session reproducibility.

In addition, to compare the water suppression performance, 2D RSI with the WET scheme and the proposed FAST water suppression scheme were measured with the same FOV and matrix size (TE/TR = 0.87ms/370ms, NA=12, FA=36°).

**Reconstruction and data analysis.** Reconstruction and post-processing: The reconstruction and data analysis were performed offline using MATLAB R2023b (MathWorks Inc., Natick, MA, USA) and LCModel[29]. The RSI data were reconstructed using the NUFFT operator[30], with a Hanning filter applied along the spatial dimensions. An adaptive coil combination was used[31].

After reconstruction, the frequency drift correction was implemented using the water scan as a reference. Residual water was removed through HLSVD[32]. Lipid signals were removed using $L_2$ regularization[23]. The lipid mask was selected as the 100 pixels with the highest lipid signal sum (LSS) (Eq 7) outside the brain per slice based on the water-suppressed MRSI data. The regularization parameter was selected in a pixel-wise manner (Figure S2.a). For each pixel, the regularization parameter was selected by increasing the parameter with an increment of one within the range of [1, 1000] until the LSS converges (Eq 8). The convergence threshold was selected as 0.005 here.

$$LSS = \sum_{i \in [0.7, 1.8] ppm} abs(S_i), \quad S: spectra \quad (7)$$

$$\frac{LSS(reg_{i+1}) - LSS(reg_i)}{LSS(reg_i)} \leq threshold \quad (8)$$

Then, the regularization parameters in the peripheral pixels (4 layers of adjacent pixels to the edge of the brain mask) were chosen to give an LSS comparable to the center pixels to avoid potential residual lipids (Eq 9).

$$LSS_{peripheral} \leqq mean(LSS_{center}) + std(LSS_{center}) \quad (9)$$

The processed spectra were quantified using LCModel[29] within the 1.6-4.2 ppm range using water data as an internal quantification reference. Metabolite basis sets were simulated in MATLAB using the density matrix formalism including the spin evolution during the P10 excitation pulse. The chemical shifts and J-coupling constants were taken from Govindaraju et al[33]. The basis set contained 21 simulated metabolites, including alanine (Ala), ascorbate (Asc), aspartate (Asp), Cr, phosphocreatine (PCr), glucose (Glc), glutamine (Gln), Glu, glutathione (GSH), glycerophosphocholine (GPC), phosphocreatine (PCho), Ins, Lac, NAA, N-acetylaspartylglutamate (NAAG), Phosphoryl ethanolamine (PE), Scylla-inositol (Scyllo), taurine (Tau), GABA, glycine (Gly), serine (Ser) were used. Altogether 14 brain metabolites were quantified, and those with CRLB lower than 100% included in the generation of metabolite maps. The basis set also includes the resonances of the macromolecules MM20 and MM16 fitted within LCModel. $T_1$ relaxation correction and water content correction were implemented[34,35]. No $T_2$ relaxation correction was applied due to the use of ultra-short TE. To evaluate the spectral quality, water linewidth was estimated by Voigt line fitting using the water data, and SNR was calculated according to (Eq 10), where $S_{NAA}$ and $S_{noise}$ are the spectral signal within [1.85,2.15] ppm and [5.1,6.1] ppm. The water suppression efficiency was evaluated according to (Eq 11), where $S_{ws}$ and $S_{nws}$ are the water-suppressed and non-water-suppressed spectral signal within range [4.3, 5.1] ppm.

$$SNR = \frac{\max(real(S_{NAA}))}{std(real(S_{noise}))} \quad (10)$$

$$Water\ Suppression\ Efficiency = \frac{\sum abs(S_{nws})}{\sum abs(S_{ws})} \quad (11)$$

The anatomical images were segmented into gray matter, white matter, and cerebrospinal fluid using FSL software[36]. Nine anatomical structural regions (caudate, cerebellum, frontal lobe, insula, occipital lobe, parietal lobe, putamen, temporal lobe, thalamus) were extracted according to a standard atlas (MNI structural atlas)[37,38]. Metabolite concentrations were analyzed in different tissue types and major brain lobes, and the inter-session reproducibility was evaluated using coefficients of variance (CV%).

**Compressed sensing optimization.** For the sake of computational efficiency, the BART package[39] was used for CS, with the default ESPIRit coil combination method[40]. The CS reconstruction can be formulated as an optimization problem to reconstruct the data from under-sampled measurements leveraging the intrinsic sparsity of the reconstructed signals. Total variation (TV) regularization was applied to the three spatial dimensions in the 3D datasets (10), where $F_u$ is the non-uniform Fourier transform operator, $x$ and $y$ are the reconstructed signal and the measured signal. $\lambda_1$ is the hyper-parameter balancing between the data fidelity term and the sparsity constraint. The alternating direction method of multipliers (ADMM) optimizer[41] was used to solve the optimization problem.

$$\min_{x} \left\| F_u x - y \right\|_2 + \lambda_1 \left\| TV(x) \right\|_1 \qquad (10)$$

Acceleration rates (R) of 1 to 6 were tested, with every one of R petals sampled. The optimal hyper-parameter $\lambda_1$ was selected by iterating over a range of values between $10^{-4}$ to 1. The structural similarity index (SSIM) between the fully sampled data (R1) and under-sampled data (R2 to R6) were assessed for water image and metabolic maps (Glu, NAA, PCr+Cr (tCr), Cho+GPC (tCho)) to select the optimal parameter (Eq 11), where $L$ is the dynamic range of the image, $\mu_x$ and $\mu_y$ are the mean value of the images, $\sigma_{xy}, \sigma_x, \sigma_y$ are the covariance and variance of the images.

$$\mathrm{SSIM}(x,y) = \frac{\left(2\mu_x\mu_y + 10^{-4}L^2\right)\left(2\sigma_{xy} + 9*10^{-4}L^2\right)}{\left(\mu_x^2 + \mu_y^2 + 10^{-4}L^2\right)\left(\sigma_x^2 + \sigma_y^2 + 9*10^{-4}L^2\right)} \qquad (11)$$

Only metabolite acquisitions were under-sampled, considering that the water acquisition was needed for coil sensitivity map generation and can be shortened by reducing the number of spectral points in practice. The hyper-parameter was optimized on one dataset and then applied to all other datasets.

# Results

**FAST water suppression efficiency.** Figure 2a shows the simulated water suppression efficiency of the WET and FAST schemes across different $B_1^+$. The FAST water suppression scheme showed substantial improvements (figure 2c) in the range of [-50, +60] % variation in the $B_1^+$, which was sufficient to cover $B_1^+$ variances at 7T for 2D and 3D measurements (figure 2b). The water sidebands in the spectra were alleviated (figure 2d) in comparison to WET. The three quartiles of the water suppression efficiency of the WET scheme and the FAST scheme were 12.3, 20.7, 30.8 and 41.3, 94.0, 177.9 respectively.

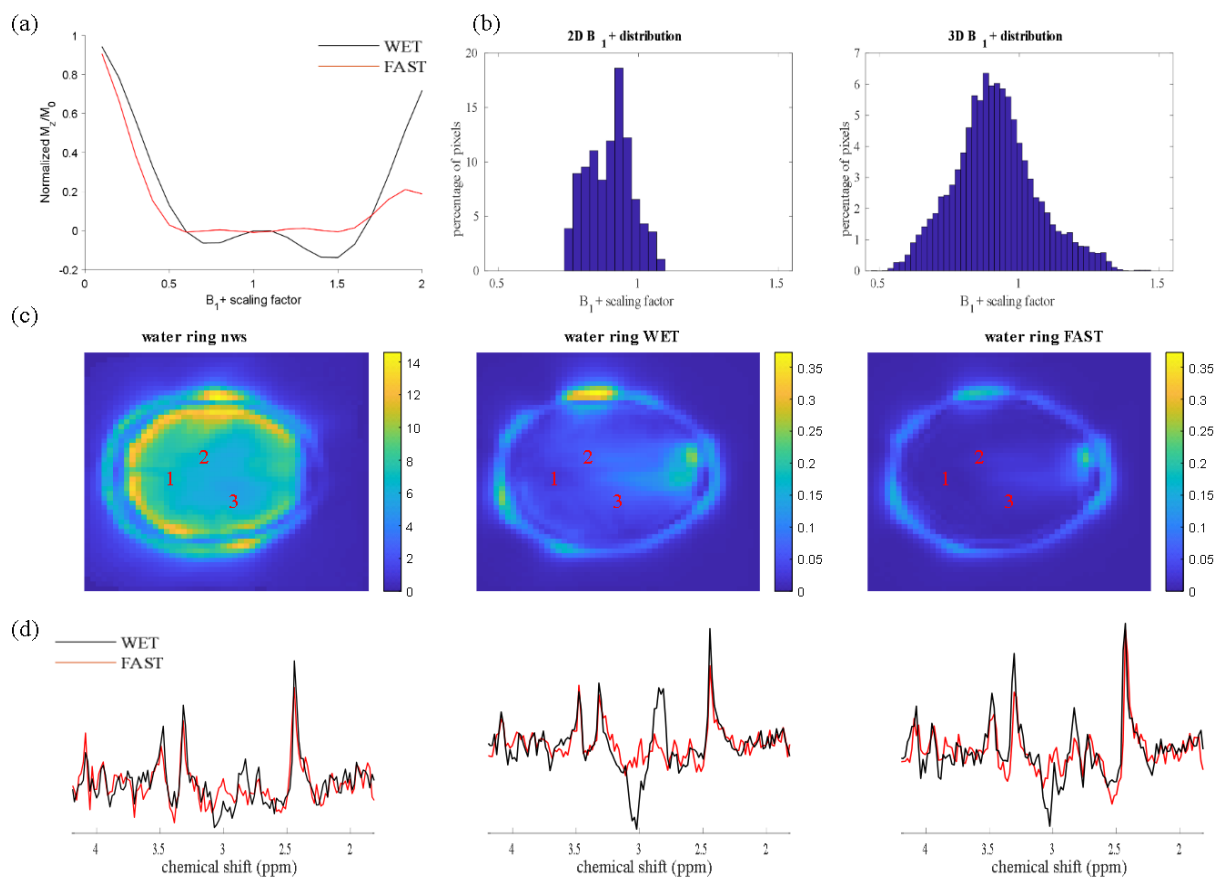

Figure 2. a) Water suppression efficiency simulation: $B_1^+ = 1$ corresponds to the nominal flip angle. b) Measured 2D and 3D $B_1^+$ distribution in a representative subject using the Siemens product tfl_rfmap sequence with the same 2D or 3D field of view as in the RSI measurements. c) Water maps (absolute spectral sum within the range of [4.3, 5.1] ppm of the water data, WET water-suppressed data, and FAST water-suppressed data; d) Sample spectra of the WET and FAST water-suppressed data, from left to right corresponding to pixel 1 to 3. The WET spectra showed obvious sidebands, which were alleviated in the FAST spectra.

**Phantom validation.** The measured metabolite concentrations in the phantom were in line with the CSI results: NAA 15.0±1.9 mM, Cr 10.0±1.1 mM, Cho 3.1±0.5 mM, Ins 14.3±2.0 mM, Glu 18.5±2.7 mM, GABA 2.2±0.6 mM for RSI and NAA 13.7±1.1 mM, Cr 10.0±0.9 mM, Cho 3.1±0.3 mM, Ins 14.4±1.1 mM, Glu 18.0±1.2 mM, GABA 2.2±0.2 mM for CSI. The metabolic maps are shown in figure S1. The discrepancy between the measured values and the ground truth values for Glu and Ins can be potentially attributed to the $T_1$ relaxation effects since a very short TR was used.

*In vivo* **validation.** Figure 3 presented 2D metabolic maps and the linewidth map. The metabolites successfully quantified over 85% of the pixels within the brain are shown (except for NAAG). The metabolic maps indicate anatomical features of gray matter (GM) and white matter (WM) for each of the metabolites. Glu, tCr have higher concentrations in the GM, tCho, Gly+Ins, and NAAG are higher in WM. The average concentrations and the third quartile CRLBs for each metabolite in WM and GM are shown in Table 1, as well as the water linewidths and SNR.

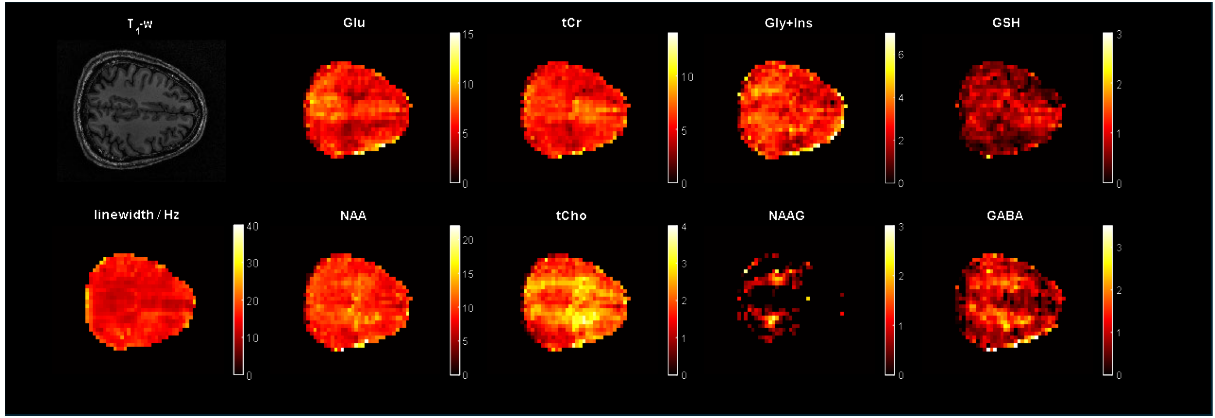

Figure 3. 2D T$_1$-w image, the water linewidth map, and 1H-FID RSI metabolic maps of brain metabolites with CRLB lower than 100% in over 85% of the brain pixels from a representative subject.

Table 1. Metabolite concentrations, third-quartile CRLBs, linewidths and spectral SNR (mean ± std)

|  |  | Glu | NAA | NAAG | Gly+Ins | tCr | tCho | GSH | GABA | Tau | LW | SNR |
|---|---|---|---|---|---|---|---|---|---|---|---|---|
| 2D | GM (conc.) | 6.3±0.4 | 8.4±0.2 | 0.6±0.1 | 2.5±0.2 | 5.7±0.2 | 1.2±0.4 | 0.5±0.02 | 0.9±0.1 | 2.6±0.2 | 14.9±0.3 | 13.6±0.5 |
|  | CRLB (75%) | 10.0±0.8 | 6.3±0.5 | 62.7±11.5 | 19.7±2.1 | 8.0±0.0 | 17.3±5.2 | 43.7±2.9 | 51.6±1.9 | 31.6±3.1 |  |  |
|  | WM (conc.) | 5.9±0.5 | 9.0±0.5 | 0.9±0.2 | 2.9±0.2 | 5.9±0.1 | 1.5±0.5 | 0.5±0.1 | 1.1±0.1 | 2.6±0.4 |  |  |
|  | CRLB (75%) | 12.7±0.2 | 6.7±0.5 | 53.7±12.7 | 19.0±2.2 | 8.7±0.5 | 13.7±4.7 | 47.7±4.5 | 45.4±3.5 | 37.7±8.8 |  |  |
| 3D | GM (conc.) | 6.2±0.5 | 8.4±0.6 | 1.0±0.2 | 2.8±0.4 | 5.6±0.1 | 1.2±0.1 | 0.7±0.1 | 1.1±0.1 | 2.6±0.4 | 26.1±1.8 | 12.0±1.6 |
|  | CRLB (75%) | 11.3±2.1 | 7.0±1.4 | 43.8±4.7 | 19.7±3.1 | 15.3±0.9 | 8.3±0.5 | 34.7±9.6 | 42.7±6.0 | 38.7±4.5 |  |  |
|  | WM (conc.) | 5.8±0.5 | 8.9±0.5 | 1.4±0.3 | 3.0±0.5 | 5.8±0.1 | 1.5±0.2 | 0.7±0.1 | 1.2±0.1 | 2.7±0.4 |  |  |
|  | CRLB (75%) | 14.3±2.6 | 7.7±0.9 | 35.3±2.9 | 20.0±3.6 | 12.7±1.2 | 9.0±0.8 | 37.7±10.8 | 44.7±6.6 | 43.0±5.9 |  |  |

Figure 4 presented 3D $T_1$ weighted anatomical images, metabolic maps and water linewidth maps. The exclusion criteria were the same as that for the 2D data. The first three slices and the last two slices were discarded due to the aliasing artifacts and the RF pulse profile, leading to altogether 15 slices. The metabolic concentrations and third-quartile CRLBs as well as the spectral quality metrics (SNR and water linewidth) are summarized in Table 1. It was observed that the central slices had better spectral quality than the peripheral slices, which was in line with the water linewidth distribution. The same GM/WM distributions were observed as that in the 2D datasets. In addition, with the same acquisition time, 3D datasets showed higher spectral SNR (15.7±0.8) than the 2D datasets in slices of comparable position as that of 2D.

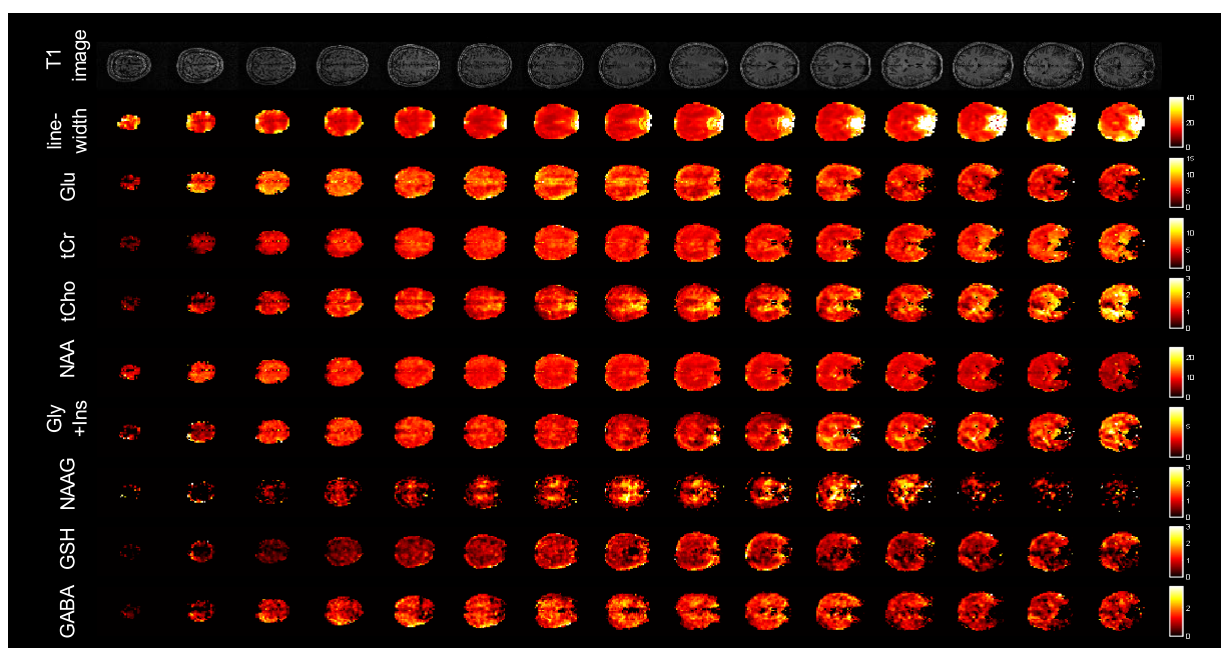

Figure 4. 3D $T_1$-w images, water linewidth maps, and $^1$H-FID RSI metabolic maps of brain metabolites with CRLB lower than 100% in over 85% of the brain pixels from a representative subject.

**Reproducibility evaluation.** Coefficients of variance were calculated in the GM, WM and several brain lobes, shown in Table 2 and 3. For 2D acquisitions, the mean CVs of NAA, tCho, tCr, Gly+Ins, and Glu were below 6% in the GM, WM, frontal lobe, and parietal lobe; the mean CVs of NAAG, GSH, and Tau were below 10%. For 3D acquisitions, the mean CVs of NAA, tCho, tCr, Gly+Ins, and Glu were below 5%, and those of NAAG, GSH, and Tau were mostly below 10% in GM, WM, frontal lobe, parietal lobe, temporal lobe, and occipital lobe. 3D data tends to show lower CVs than 2D data in the frontal lobe and parietal lobe.

Table 2. 2D inter-session CVs (%) for metabolites determined in different tissues and brain lobes.

|  | Glu | NAA | NAAG | Gly+Ins | tCr | tCho | GSH | GABA | Tau |
|---|---|---|---|---|---|---|---|---|---|
| GM | 2.3±1.8 | 0.8±0.4 | 8.4±7.9 | 2.7±3.1 | 3.5±1.4 | 4.6±5.9 | 6.9±4.8 | 2.6±1.6 | 4.4±4.3 |
| WM | 4.3±1.9 | 2.1±1.1 | 5.9±5.6 | 1.4±1.9 | 4.4±1.2 | 4.2±2.6 | 9.3±7.0 | 4.2±2.2 | 7.5±4.3 |
| FRONTAL | 3.8±1.4 | 1.0±0.4 | 8.6±4.4 | 2.5±2.1 | 4.2±0.3 | 5.9±2.7 | 6.6±3.2 | 3.2±0.7 | 6.0±4.8 |
| PARIETAL | 2.9±1.8 | 1.2±1.2 | 3.0±3.3 | 3.9±3.4 | 3.4±1.2 | 3.9±1.7 | 12.8±4.1 | 4.7±1.2 | 3.8±2.7 |

Table 3. 3D inter-session CVs (%) for metabolites determined in different tissues and brain lobes.

|  | Glu | NAA | NAAG | Gly+Ins | tCr | tCho | GSH | GABA | Tau |
|---|---|---|---|---|---|---|---|---|---|
| GM | 0.9±0.3 | 0.7±0.4 | 2.4±2.5 | 2.6±0.9 | 0.7±0.1 | 1.5±1.4 | 1.5±0.3 | 1.1±0.7 | 2.5±0.8 |
| WM | 0.4±0.3 | 0.3±0.2 | 1.6±1.4 | 2.4±1.1 | 1.4±0.4 | 2.1±0.5 | 1.9±1.1 | 1.2±0.7 | 2.5±0.8 |
| FRONTAL | 0.6±0.2 | 0.4±0.2 | 1.6±0.4 | 1.0±0.7 | 1.3±0.6 | 1.0±0.5 | 1.1±0.6 | 0.8±0.8 | 2.3±1.0 |
| PARIETAL | 0.9±0.8 | 1.0±0.1 | 2.3±1.2 | 3.9±2.2 | 2.0±0.4 | 1.6±2.0 | 1.4±1.1 | 1.4±0.8 | 2.7±1.6 |
| *TEMPORAL* | 3.5±1.2 | 1.3±0.8 | 10.2±3.5 | 3.2±2.0 | 1.1±1.0 | 2.0±2.0 | 3.5±2.2 | 4.2±3.2 | 1.8±0.6 |
| *OCCIPITAL* | 3.9±1.4 | 1.2±1.0 | 8.6±6.5 | 2.7±0.4 | 0.8±0.5 | 4.6±2.6 | 8.0±2.3 | 2.2±1.5 | 7.0±0.6 |

**Retrospective Compressed Sensing Acceleration.** The optimization process showed that $\lambda_1 = 0.05$ gave the highest SSIM in terms of the water images and metabolic maps concerning the fully sampled data. Thus, $\lambda_1 = 0.05$ was used for all datasets and acceleration rates.

Figure 5 shows the accelerated 3D metabolic maps with acceleration rates (R) of 1, 2, 3, 4, 5, and 6 from the central five slices. It is shown that for Glu, the GM/WM contrast was preserved until R=3. The contrasts started to change with increasing acceleration rates. The mean concentrations, CRLBs, SSIM, and spectral SNR for different acceleration rates were shown in Table 5. Spectral SNR and SSIM decreased as R increased, with an average SSIM larger than 0.8 until R=3.

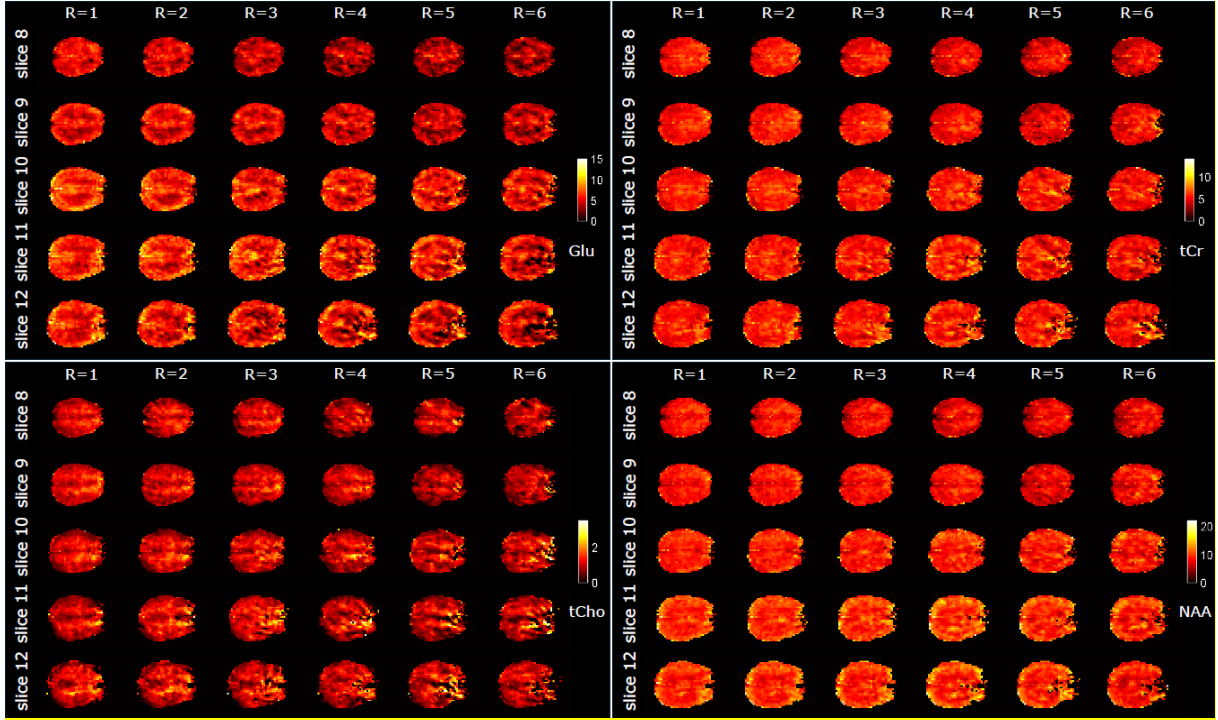

Figure 5. 3D Glu, tCr, tCho, NAA metabolic maps at different acceleration rates R = 1, 2, 3, 4, 5, 6 from the central five slices in a representative subject.

Table 4. Metabolite concentrations, third-quartile CRLBs, and SSIM at different acceleration rates (mean ± std)

| | | Glu | NAA | tCho | tCr | Gly+Ins | SNR |
|---|---|---|---|---|---|---|---|
| *R1* | Conc | 6.2±0.1 | 9.6±0.2 | 1.3±0.2 | 6.1±0.4 | 3.3±0.5 | 12.2±1.8 |
| | CRLB (75%) | 12.7±0.5 | 6.7±0.9 | 14.3±0.9 | 8.3±0.5 | 16.7±3.1 | |
| | SSIM | 1 | 1 | 1 | 1 | 1 | |
| *R2* | Conc | 5.7±0.1 | 9.5±0.3 | 1.3±0.2 | 6.1±0.3 | 3.1±0.5 | 11.1±0.9 |
| | CRLB (75%) | 13.7±0.5 | 6.7±0.9 | 15.7±1.2 | 8.3±0.5 | 18.3±3.4 | |
| | SSIM | 0.88±0.03 | 0.89±0.04 | 0.87±0.02 | 0.90±0.04 | 0.85±0.04 | |
| *R3* | Conc | 5.1±0.3 | 9.2±0.7 | 1.2±0.1 | 5.7±0.2 | 2.9±0.4 | 10.2±0.6 |
| | CRLB (75%) | 16±1.4 | 7.3±1.2 | 17.3±0.5 | 9.3±0.5 | 19.7±3.1 | |
| | SSIM | 0.83±0.03 | 0.84±0.04 | 0.81±0.02 | 0.86±0.04 | 0.79±0.04 | |
| *R4* | Conc | 4.9±0.6 | 8.3±1.2 | 1.1±0.1 | 4.9±0.4 | 2.8±0.4 | 9.1±0.7 |
| | CRLB (75%) | 17.3±2.6 | 7.7±1.7 | 19.7±0.5 | 10.0±1.4 | 19.7±3.3 | |
| | SSIM | 0.80±0.04 | 0.81±0.06 | 0.78±0.03 | 0.82±0.05 | 0.76±0.04 | |
| *R5* | Conc | 4.4±1.0 | 8.3±1.8 | 1.0±0.1 | 4.7±0.6 | 2.9±0.4 | 8.2±0.6 |
| | CRLB (75%) | 19.0±4.2 | 10.3±4.7 | 20.7±2.4 | 12.3±4.0 | 21.0±2.9 | |
| | SSIM | 0.77±0.04 | 0.78±0.06 | 0.75±0.02 | 0.80±0.05 | 0.74±0.04 | |
| *R6* | Conc | 4.4±1.4 | 8.4±2.6 | 1.0±0.2 | 4.7±0.9 | 2.9±0.4 | 7.9±0.5 |
| | CRLB (75%) | 20.7±4.5 | 11.0±5.0 | 23.3±2.5 | 14.0±4.3 | 22.0±4.1 | |
| | SSIM | 0.74±0.04 | 0.75±0.06 | 0.72±0.03 | 0.76±0.05 | 0.71±0.04 | |

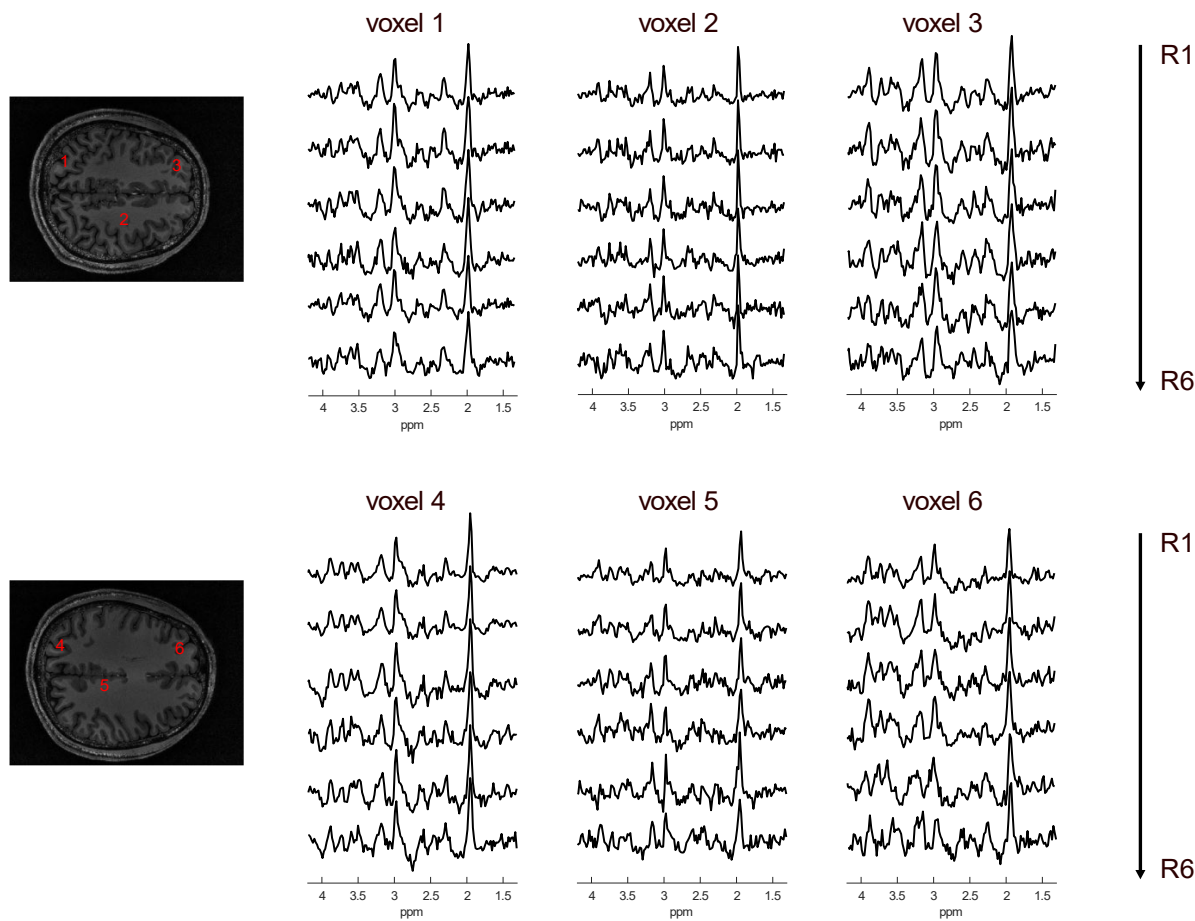

Figure 6. Representative spectra from six different positions in two different slices at acceleration rates R = 1, 2, 3, 4, 5, 6 from a representative subject.

## Discussion

In this study, we developed a 2D and 3D whole-brain non-lipid-suppression ultra-short TE MRSI acquisition scheme using the rosette trajectory. The method showed excellent inter-session reproducibility with an optimized water suppression scheme and processing pipeline. Additionally, it is shown that with CS, the 3D acquisition had the potential to be further shortened to around 2min.

Residual water and lipid signals are two main sources of spectral contamination for $^1$H MRSI. The sidebands of the water resonance can fall inside the metabolite range, and overlap with some key metabolites such as glutamate, causing bias or even failure in metabolite quantification. With the FAST water suppression scheme, the water suppression efficiency increased across a wide range of $B_1^+$, which alleviated water sidebands and improved spectral quality. The short total duration of the FAST scheme enables the use of a short TR, further reducing the overall scanning time.

Lipids, another intensive signal from the skull, can contaminate the spectra due to signal propagation through the point spread function and the imperfection of the excitation pulse profile. Inversion recovery (IR) and outer volume suppression (OVS) are two common methods of lipid suppression. However, the additional pulses can increase RF energy deposition, leading to the use of a longer TR and, thus, longer acquisition time. Whole-brain MRSI is usually applied without OVS to obtain metabolic information in the peripheral cortical regions. Such acquisition often relies on postprocessing strategies to suppress lipid signals such as data extrapolation[24], $L_2$ regularization[23], Fast LIpid reconstruction and removal Processing (FLIP)[25], etc, with $L_2$ regularization being the most commonly applied approach. The $L_2$ regularization was directly applied to the dataset first, however, it was challenging to find a regularization parameter that could effectively suppress lipid signals in the peripheral voxels without over-suppressing metabolite signals in the central voxels. Note that with a regularization parameter that is optimal for the center voxels, residual lipids may remain in the peripheral voxels, potentially affecting metabolite quantification, particularly for NAA. Thus, we further optimized the $L_2$ regularization method to have different regularization parameters in different pixels, where the peripheral pixels tend to have higher regularizations. This is coherent with the intuition that the peripheral pixels have higher lipid contamination and thus need heavier regularizations. In addition to the $L_2$ and pixel-wise $L_2$ regularization methods that were applied in the current study, the data extrapolation[24] method and the FLIP[25] method might also be applied. However, these methods may require further adaptation, as their performance was optimized using data acquired with Cartesian trajectories at a TE of 16 ms or non-Cartesian trajectories at a TE of 50ms. In those cases, lipid signals decayed around 3.3 or 50 times more than with the ultra-short TE of 0.83/0.87ms used in this study. Additionally, Cartesian trajectories typically offer better point spread functions and less lipid signal propagation compared to the non-Cartesian ones, unless dedicatedly designed[6].

We observed strong GM/WM contrasts in metabolic maps, with higher levels of tCho and NAAG in the WM, and higher Glu levels in the GM. The spatial patterns of their concentrations were consistent with previous results[28,42]. Similar to other studies, the NAA concentration was relatively uniform based on Figure 3-5, though the values were slightly higher in the WM in the 2D slice, which can be potentially biased by the inhomogeneous spectral quality such as SNR and linewidth[43–45] across voxels. Using basis sets with a matched linewidth to the *in vivo* data could potentially alleviate the quantification bias[46]. For GSH and GABA, which were shown to be higher in the GM,, no obvious contrasts were found. This can be due to their low concentrations and high spectral overlapping with intensive metabolites[20,42]. Dedicated methods such as MEGA[38] and SLOW[39] editing will be needed for a robust quantification of these metabolites.

The inter-session reproducibility of the 2D and 3D sequences were evaluated. Overall, the CVs of NAA, tCr, Gly+Ins, tCho, and Glu ranged from 0.2-3.7%, 1.4-5.5%, 0.5-8.7%, 1.3-9.7%, 0.1-6.7% and 0.1-2.6%, 0.3-2.5%, 0.1-6.8%, 0.1-7.5%, 0.1-5.6% for 2D and 3D respectively. Comparing the CVs to previous studies can be difficult due to the different field strengths, quantification references, and ROIs selected. At 7T, using water as an internal quantification reference, average CVs of 3-7%, 3-6%, 4-7%, 3-5%, 3-7% for NAA, tCr, Ins, tCho and Glu in 3D data were reported[13]; using tCr as the quantification reference, CVs of 3–8%, 2–10%, 5–13%, 6–16% and 7.0%, 5.9%, 7.0%, 8.1% for NAA/tCr,, tCho/tCr, Ins/tCr, Glu/tCr were reported respectively[50,51]. At 9.4T, mean CVs reported are ranging from 6.2-8.5%, 4.6-5.1%, 6.5-10.3%, 7.4-16.7% for NAA/tCr, tCho/tCr, Glu/tCr, and Ins/tCr[42]. The present CV values were similar to but slightly lower than those reported in these studies. Note that the 3D sequence tends to have lower CVs than 2D acquisition, likely due to the improved SNR typically achieved with 3D acquisitions. The CVs were, in general, higher in the frontal lobe, which can be attributed to the higher $B_0$ field inhomogeneity since the frontal lobe is close to the air-filled sinuses. Higher order shims and a local shim coil can be incorporated in future experimental setups to improve $B_0$ field inhomogeneity, especially at the frontal lobe, by static and dynamic shimming[52,53]. Except for the SNR and $B_0$ inhomogeneity, the reproducibility can also be influenced by $B_1^+$ inhomogeneities and subject motions. Incorporating navigators in the future could reduce motion artifacts. The acquired $B_1^+$ maps could also be included in the correction of metabolite quantification.

We explored the possibilities of combining rosette spectroscopic imaging with compressed sensing. It was shown that for 3D data, the SSIM index remained above 0.85 and 0.80 for Glu, NAA, tCr, and tCho for all subjects until R=2 and R=3, corresponding to an acquisition time of 2:15min and 1:50 min, respectively. It was noted that different metabolites might have different maximum acceleration rates[54]. For example, the SSIM for Glu, NAA, tCr, and tCho were above 0.8 until R=3, while until R=2 for Gly+Ins. Furthermore, with the increase of acceleration rates, the metabolite concentrations decreased. This might be related to the degradation in SNR with a higher acceleration factor, leading to bias in metabolite quantification[43,44]. Note that petals were under-sampled uniformly, i.e., every R petal was sampled. This is because the rosette trajectory was designed based on the stochastic trajectory[16], which is intrinsically incoherent. Different under-sampling schemes including regular sampling and random sampling of the petals were evaluated at R=2 to explore the influence of under-sampling strategies. It was shown that the metabolic maps obtained with randomly sampled petals gave lower SSIM with respect to the fully sampled data than those of regularly sampled (figure S3). This was coherent with the simulation results, where randomly sampled trajectories had higher average sidelobe level and maximum sidelobe level, indicating lower trajectory incoherence (table S1)[55].

This study has several limitations. First, compared with conventional CSI, fast trajectories are more prone to the $B_0$ field inhomogeneity, which can cause a deviation of the actual trajectory from the designed trajectory, leading to a worse point spread function and thus broader linewidth. To further improve the spectral quality, $B_0$ field correction can be included. Besides, as a proof-of-concept, the BART package and the total variation regularization were used for compressed sensing, and the same regularization parameter was used for different acceleration rates. However, total variation regularization can lead to smooth reconstructions, which can blur brain structures[56]. Different regularization methods such as spatial-spectral regularization to enforce x-t sparsity[12,57] can be explored, and the regularization parameters can be further fine-tuned in future studies.

# Conclusion

2D and 3D whole-brain metabolic maps of major [1]H metabolites such as NAA, Glu, tCr, tCho, and Gly+Ins, were achieved using the non-lipid-suppressed ultra-short-TE-FID RSI with improved water suppression performance at 7T. The acquisition took 20 seconds to acquire a one-shot 2D slice with a resolution of 4.48x4.48mm$^2$ and 5:30min to acquire 3D metabolic maps with a resolution of 4.48x4.48x4.50 mm$^3$. With compressed sensing, the 3D acquisition has the potential to be further reduced to around 2min. The current implementation of [1]H-FID-RSI allows for fast and reproducible measurement of the spatial metabolic distribution in the human brain, thereby paving the way for a better understanding of brain functions and pathology.

# Acknowledgements

This work is supported by Swiss National Science Foundation NO.: 189064, 213769. We acknowledge access to the facilities and expertise of the CIBM Center for Biomedical Imaging, a Swiss research center of excellence founded and supported by Lausanne University Hospital (CHUV), University of Lausanne (UNIL), Ecole Polytechnique Federale de Lausanne (EPFL), University of Geneva (UNIGE), and Geneva University Hospitals (HUG).

# Conflict of Interest Statement

The authors state that they have no known financial conflicts of interest or personal relationships that could have influenced the work presented in this paper.

# Supplementary Materials

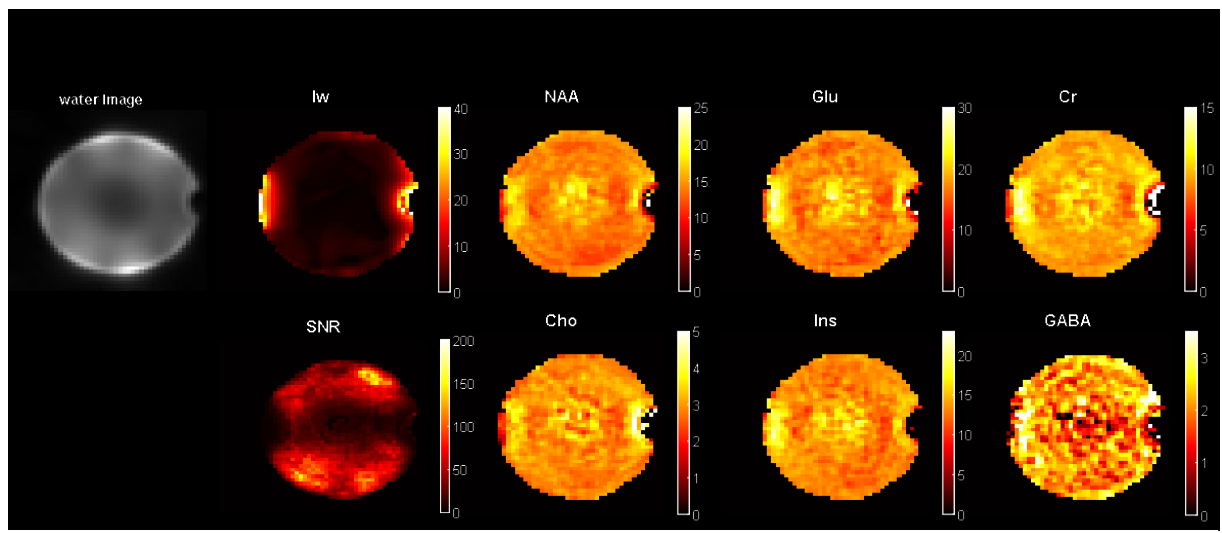

Figure S1. Water image, water linewidth map, SNR map, and $^1$H-FID RSI metabolic maps of phantom metabolites acquired with a spatial resolution of 4.48x4.48mm$^2$ in 30 seconds. 75% of the phantom pixels have CRLB lower than 4%, 5%, 5%, 5%, 6%, and 29% for NAA, Cr, Ins, Glu, Cho, and GABA, respectively.

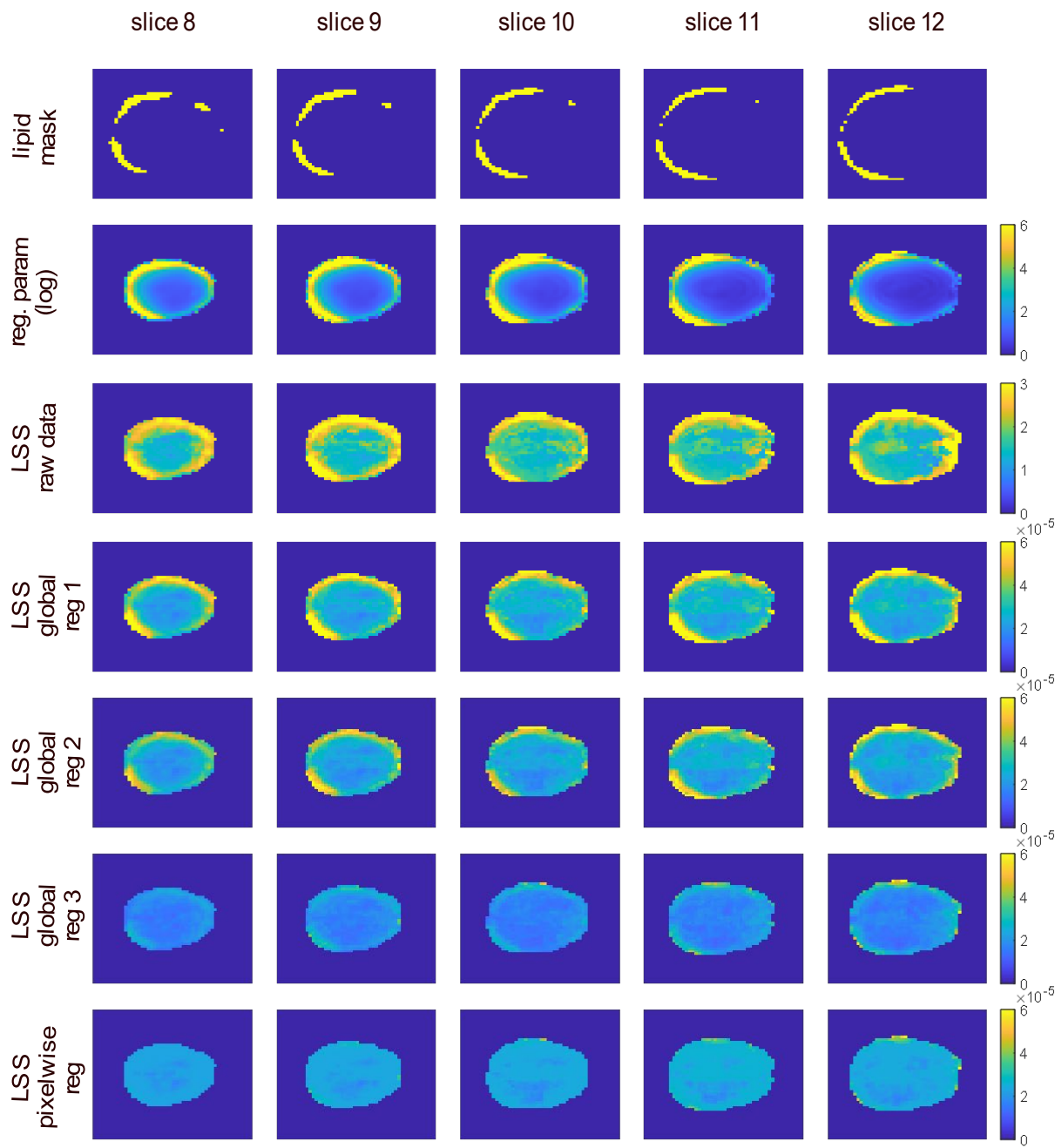

Figure S2. Pixel-wise $L_2$ regularization. Row 1: the selected lipid mask; row 2: the final regularization parameter selected by the two-step pixel-wise $L_2$ method in log scale; row 3: the lipid signal sum of the water-suppressed data after HLSVD and before lipid removal; row 4-6: the lipid signal sum of the data after the original $L_2$ lipid removal method with regularization parameter being 25, 45, and 700, which corresponded to the average regularization parameter in the center brain region, peripheral brain region given in the first step and the average regularization parameter in the peripheral brain region given in the second step respectively; row 7: the lipid signal sum of the data after the pixel-wise $L_2$ lipid removal method with the regularization parameters shown in row 2.

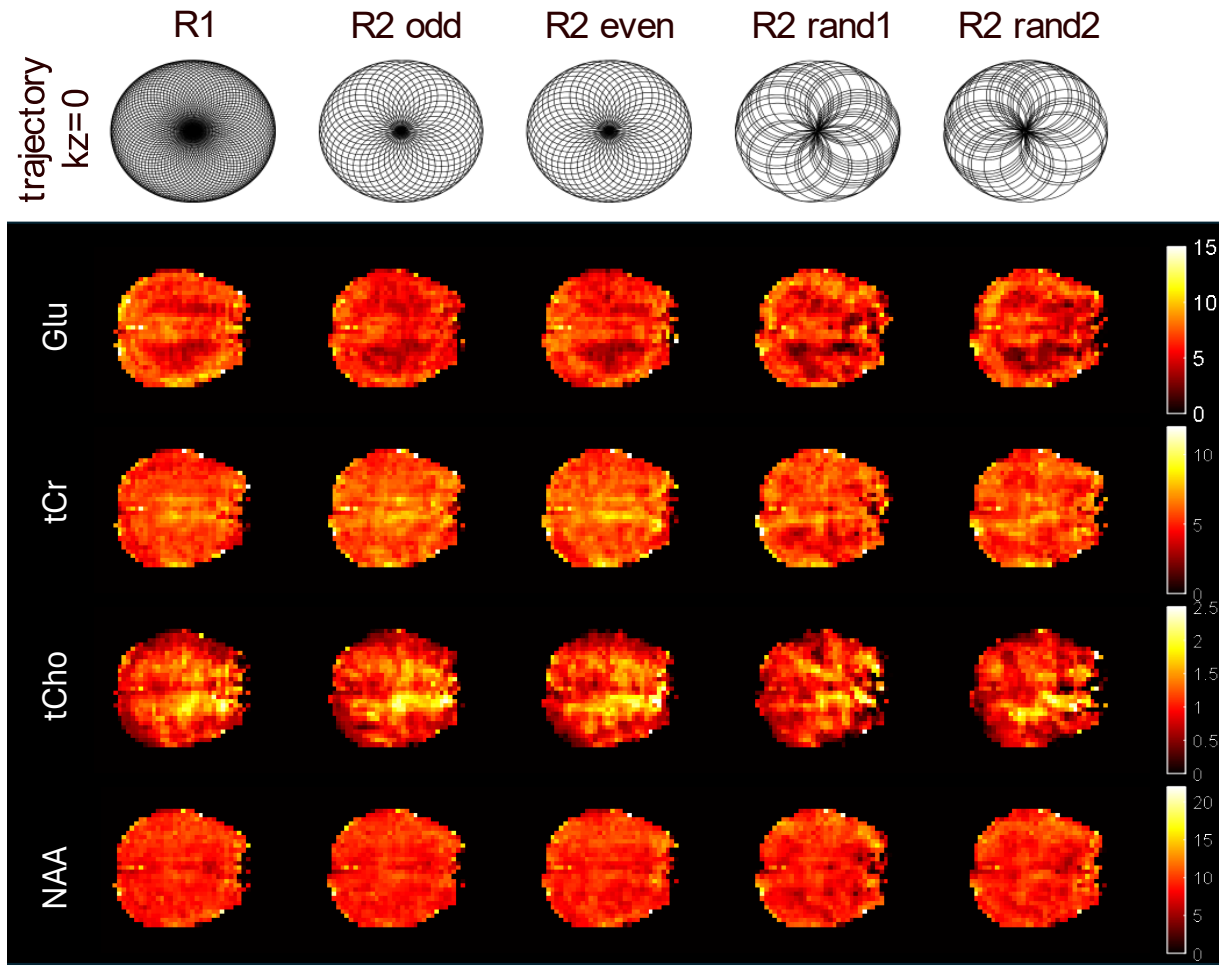

Figure S3. Different under-sampling schemes: a) k-space trajectories; b) metabolic maps reconstructed from corresponding trajectories at acceleration rate R = 2.

Table S1. The SSIM and incoherence metrics of the corresponding under-sampling schemes shown in figure S3. ASL and MSL refer to Average Side Lobe amplitude and Maximum Side Lobe amplitude.

|  | ASL | MSL | SSIM | | | | CRLB (75%) | | | |
|---|---|---|---|---|---|---|---|---|---|---|
|  |  |  | Glu | tCr | NAA | tCho | Glu | tCr | NAA | tCho |
| R1 | 0.0006 | 0.3186 | 1 | 1 | 1 | 1 | 9 | 7 | 6 | 13 |
| R2-even | 0.0117 | 0.3530 | 0.91 | 0.91 | 0.87 | 0.92 | 11 | 7 | 6 | 15 |
| R2-odd | 0.0118 | 0.3497 | 0.91 | 0.91 | 0.85 | 0.91 | 11 | 7 | 6 | 16 |
| R2-rand1 | 0.0154 | 0.3955 | 0.87 | 0.83 | 0.78 | 0.85 | 13 | 9 | 7 | 17 |
| R2-rand2 | 0.0151 | 0.3790 | 0.87 | 0.85 | 0.80 | 0.88 | 13 | 8 | 6 | 17 |